\documentclass[useAMS,usenatbib]{mn2e}
\usepackage{graphicx,graphics,amsmath}
\usepackage{latexsym}

\usepackage{float}
\usepackage[caption = false]{subfig}

\onecolumn
\usepackage{xcolor}

\begin{document}

\title[Max mass of HS from shock induced PT in cold NS]{Maximum mass of hybrid star formed via shock induced phase transition in cold neutron stars}

\author[ Ritam Mallick \& Shailendra Singh \& Rana Nandi]
{ Ritam Mallick$^{1}$\thanks{mallick@iiserb.ac.in},Shailendra Singh$^{1}$\thanks{shailendra17@iiserb.ac.in}, Rana Nandi$^{2}$\thanks{}\\
	$^{1}$ Indian Institute of Science Education and Research Bhopal, Bhopal, India \\
	$^{2}$Polba Mahavidhyalaya, Hoogly, West Bengal 712148, India }

\maketitle
\begin{abstract}
This article studies the maximum mass limit of the quark star formed after the shock-induced phase transition of a cold neutron star. By employing hadronic and quark equation of state that satisfies the current mass bound, we use combustion adiabat conditions to find such a limit. The combustion adiabat condition results in a local or a global maximum pressure at an intermediate density range. The maximum pressure corresponds to a local or global maximum mass for the phase transformed hybrid star. The phase transition is usually exothermic if we have a local maximum mass. The criteria for exothermic or endothermic phase transition depend on whether the quark pressure/energy ratios to nuclear pressure/energy are smaller or greater than 1. We find that exothermic phase transition in a cold neutron star usually results in hybrid stars whose mass is smaller than a parent neutron star.
The phase transition is endothermic for a global maximum pressure; thereby, one gets a global maximum mass. Hybrid stars much massive than phase transformed local maximum mass can be formed, provided there is some external energy source during the phase transition process. However, for some cases, even massive hybrid stars can form with exothermic phase transition for EoSs having global maximum pressure.
\end{abstract}

\section{Introduction}

One of the most exciting question in recent astrophysics is that whether quark matter (QM) exists at the neutron star (NS) interiors. The strongly interacting matter theory predicts that at high density and temperature, the ground state of matter consists of quarks and gluons \citep{shuryak}. It means that the ordinary hadronic matter (HM) would undergo a deconfinement transition to a new quarks and gluons phase under such extreme conditions. The deconfinement transition from the hadron to quark-gluon phase has been observed at high temperature and small baryon density in ultra-relativistic heavy-ion collision \citep{gyulassy,andronic}. However, whether the quark phase exists inside NS cores (where density is several times nuclear saturation density) is still an open question.

The cores of NSs cannot be directly observed; therefore, we need to model NS matter from crust to the core to obtain the properties of matter at the core. Although the crust and the NS's outer surface can be modeled quite accurately, the lack of exact theory for the strongly interacting matter at densities beyond saturation density makes it difficult to model NS cores precisely. However, this limitation can be overcome by integrating astrophysical observation of NS mass and radius and the gravitational wave observation of binary NS mergers (BNSM).

The most recent and major breakthrough came in the previous decade with the precise measurement of two massive NS masses \citep{demorest,antonidis,cromartie}. However, to put more severe constraints on the equation of state (EoS), subsequent precise measurement of both mass and radius is needed. The uncertainty in the NS radius measurement is still quite large. Neutron Star Interior Composition Explorer (NICER) was launched to address this problem, aiming to measure mass and radius with 5\% accuracy. Recently it estimated the mass and radius of PSR J0030+0451 (R=$12.71^{+1.14}_{-1.19}$ and M=$1.34^{+0.15}_{-0.16}$) \citep{riley,miller}.

On the other hand, a significant breakthrough also came from the gravitational wave detection of GW170817 \citep{abbott}. The merger's inspiral phase induces the high tidal deformability in the stars due to their strong gravitational fields \citep{hinderer}. As the tidal deformability is strongly dependent on the NS radius ($\Lambda \sim R^5$), it puts a constraint on the radius and subsequently on the EoS \citep{annala,most,rana,zhang}. Analyzing GW170817, the tidal deformability bound was found to be $\Lambda \le 600$ \citep{abbott2}. It constrains the 1.4 solar mass NS should have a radius between 12.9-13.5 km.
All these recent observations have put severe constrain on the EoS, and many of the EoS has been discarded; however, whether QM can occur inside NS or not is still not settled. A recent study by Annala et al. \citep{annala} suggested that it is more likely that QM should exist inside NSs. Other investigations \citep{rana} also argued that QM's existence is possible in NSs interiors.

If QM does exist, it is likely to be generated during supernova explosion or in the hypermassive NS after the merging of the binary NSs. However, the phase transition (PT) of hadron to QM in NS interior can also occur in cold NS if there is some density fluctuation at their interior during glitch or mass accretion \citep{chubarian, glen,burderi}. If the PT does happen in cold NSs, it is likely to be a shock-induced PT \citep{bhattacharyya,prasad1}. The shock-induced PT has been studied in considerable detail in the literature. 
There are two aspects of such studies; one is the kinematic calculation, which employs the Rankine-Huguenot jump condition to classify the PT process \citep{cho,bhattacharyya,drago2007,igor,singh} and the other aspect is the dynamics of the PT where time evolution of the combustion of HM to QM is studied \citep{olinto,lin,abdi,prasad1,prasad2,ss1}. In this article, we will follow the kinematic approach. In the literature there are two very different PT scenarios: a) the PT propagates as a slow combustion (deflagration) \citep{drago2007,singh} b) the PT is a fast detonation \citep{bhattacharyya,bhat2,igor}. There are also studies that the PT can start as a deflagration, but instabilities (generated due to gravity) could quickly turn it into a fast detonation \citep{herzog,horvath}. In most of this calculation,
the dynamical evolution of the shock wave was not considered,
and the combustion process was categorized by employing the
energy-momentum and baryon number conservation across the
shock discontinuity.

In a recent work \citep{irfan-mnras} the Combustion adiabat (CA) or the Chapman-Jouget adiabat equation were employed to study the PT's kinematics. The study suggested a maximum quark pressure reflected in the maximum mass of the resulting hybrid star (HS) (a star with quark core and HM outer layers)  from PT. This work employs different EoSs, consistent with the recent astronomical constraints, and study the PT in NSs. In this work, we study whether the PT happening in a cold NS is exothermic or endothermic, and their implication on the maximum mass of a phase transformed HS. If the energy output of such conversion is exothermic, then the PT is a viable process in such cold NSs, but if it is endothermic, some external energy source is needed for the PT to occur. 

\section{Methodology}

The Rankine-Huguenot (RH) condition connects the matter quantities across a shock discontinuity, conserving flux and energy-momentum \citep{landau,irfan-mnras}. The RH equations can be solved to get a single equation relating matter quantities across the discontinuity known as Taub adiabat \citep{taub,thorne}. If the initial and final state (EoS) across the discontinuity is different due to chemical energy changes, the relation is known as CA.
The CA is easier to solve than the RH equations as it is a single equation involving matter quantities and not involving the matter velocities. The velocities can be obtained after solving the CA equation.

The RH condition for a shock discontinuity at $y=z=0$ and perpendicular to the $x$-axis are given by 
\begin{align}
 w_h\gamma_h^2v_h=w_q\gamma_q^2v_q ;\\
 w_h\gamma_h^2v_h^2+p_h=w_q\gamma_q^2v_q^2+p_q ;\\
 n_hv_h\gamma_h=n_qv_q\gamma_q\equiv j ;
\end{align}
with $\gamma$ is the Lorentz factor, $w$ the enthalpy ($w=\epsilon+p$) and $u^{\mu}=(\gamma,\gamma v)$ is the normalized 4-velocity of the fluid. Assuming the PT happening at the dicontinuity separating the two phases (can be thought of unburnt or upstream matter and burnt or downstream matter). In this study the unburnt matter is the HM (denoted by $'h'$) and QM being the burnt matter (denoted by $'q'$).

We assume that the PT happens as a single discontinuity, and the front propagates, separating the two phases. 
Therefore we denote $'h'$ as the initial state ahead of the shock front; in our analysis, it is HM, and $'q'$ as the final state behind the shock, the final burned QM.

The CA equation can be obtained by solving the above three equation (for detail please refer \citep{irfan-mnras}) and is given by
\begin{equation}
 {\mu_q}^2 - {\mu_h}^2 = (p_q - p_h)\left(\frac{\mu_h}{n_h} + \frac{\mu_q}{n_q}\right).
\end{equation}

Another useful way of writing the CA equation is in terms of a quantity $X$ (defined as $X_i=w_i/n_i^2=\mu_i/n_i$) and the CA equation then takes the form
\begin{equation}
 w_qX_q - w_hX_h=(p_q - p_h)(X_h + X_q).
\end{equation}

Knowing the unburnt matter properties (input values for the HM), we can solve for the burnt phase quantities (pressure, density, chemical potential) if the EoSs are known. In the $X,p$ diagram we can plot the curves for the HM as well as QM and due to the difference in chemical energy we obtain two different curve. The line that connects a point in the HM curve (a given initial state) and the corresponding point (obtained by solving the CA equation) in the QM curve (the final state) is called the ``Rayleigh line'' (RL). Once the matter properties are known the velocity of the corresponding phases are given by \citep{thorne} 
\begin{align}
 |v_h|=\left[\frac{(p_q-p_{ h}) (\varepsilon_q+p_h)}{(\varepsilon_q-\varepsilon_h) (\varepsilon_n+p_q)}\right]^{1/2}, \nonumber \\
~~|v_q|=\left[\frac{(p_q-p_{ h}) (\varepsilon_h+p_q)}{(\varepsilon_q-\varepsilon_h) (\varepsilon_q+p_{ h})}\right]^{1/2}.
\label{evel}
\end{align}

To solve the CA equations, we need the EoSs of the HM and QM.
For the EoS of the HM we consider three 
relativistic mean-field (RMF) models:
S271v2 \citep{Horowitz:2002mb}, FSUGarnet \citep{Utama:2016tcl} 
and DD2 \citep{Typel:2009sy} and a non-relativistic
Skyrme-type model SkI3 \citep{Reinhard:1995zz}.
 At low density, the Baym-Pethick-Sutherland EoS \citep{Baym:1971pw}
of the crust is added to all of these hadronic EoS in a
thermodynamic consistent fashion.

\begin{figure*}
\centering

   \includegraphics[width = 3.45in,height=2.5in]{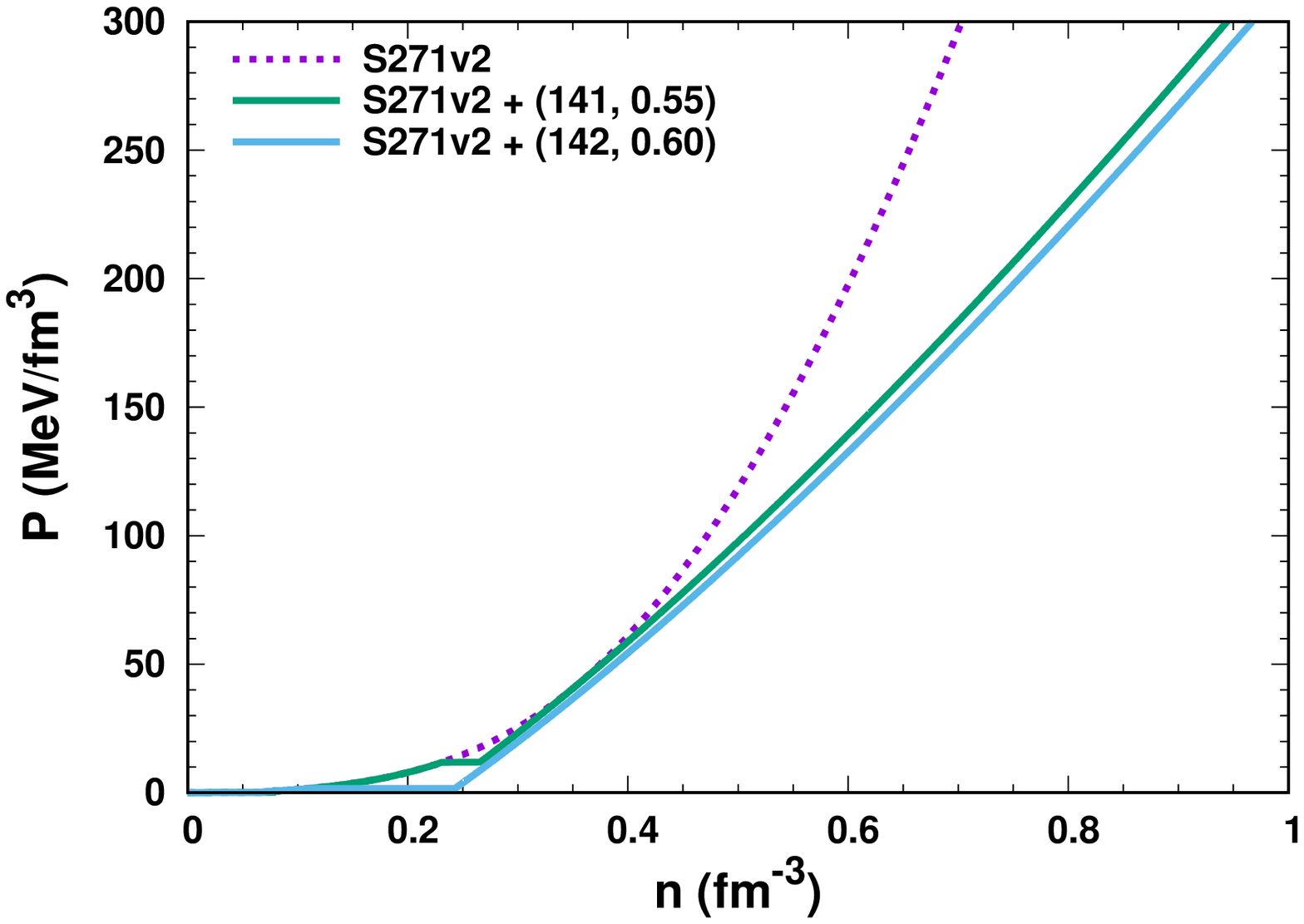}
  \includegraphics[width = 3.45in,height=2.5in]{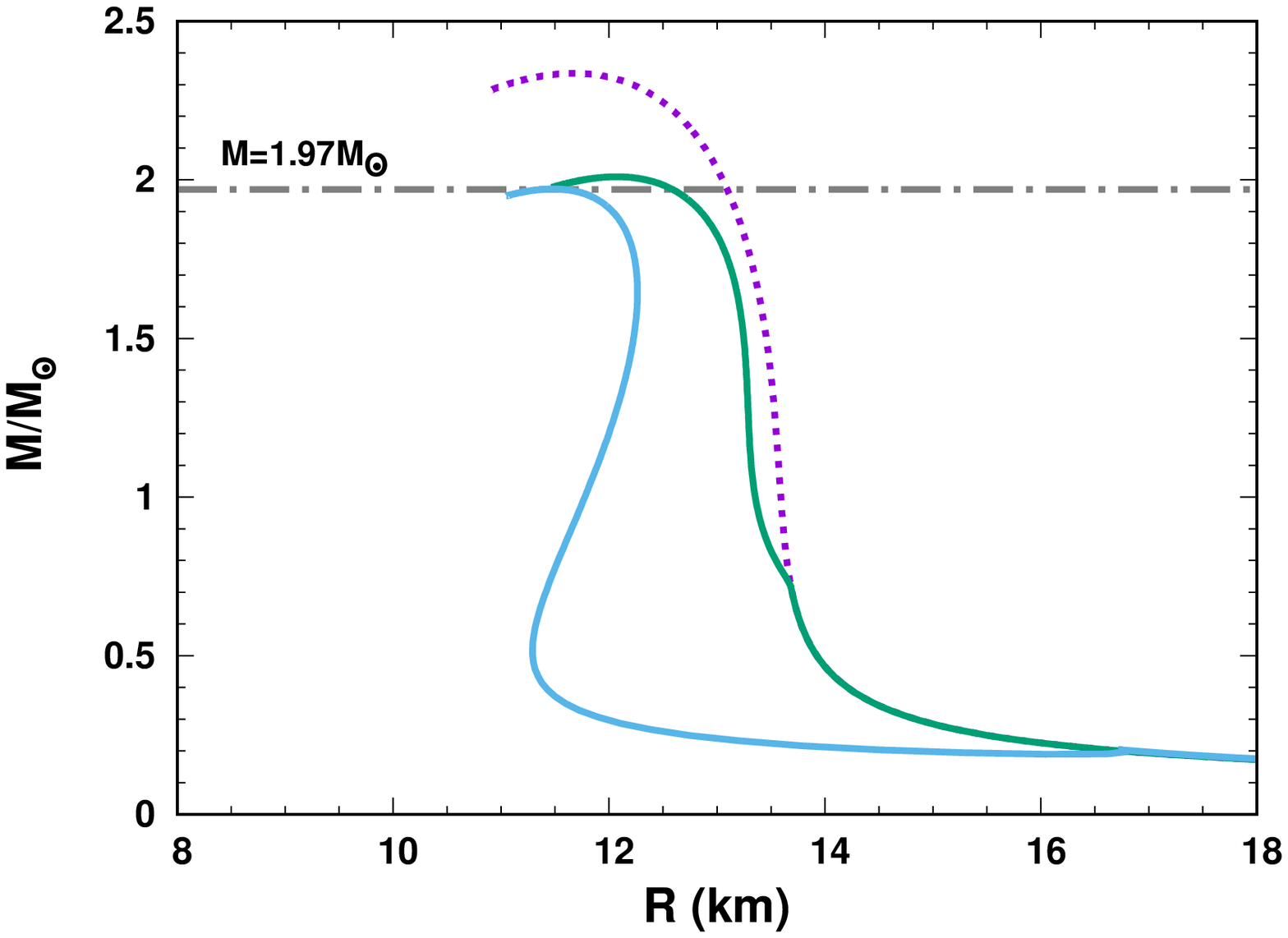}
  
  \hspace{0.5cm} \scriptsize{(a)} \hspace{8.5 cm} \scriptsize{(b)} 

 \caption{Equations of state (left panel) and mass-radius relations (right panel) for S271v2 hadronic model. Dotted curves are for pure hadronic star and  solid curves  
 represent HS where the quark part is described by different combinations of bag model parameters
 ($B_{\rm eff}^{1/4}, a_4$). The horizontal dash-dotted
 line represents the lower bound on maximum mass ($M_{\rm max}=1.97M_\odot$).}
 \label{fig:eos}
\end{figure*}

The EoS of the QM is constructed by adopting
the modified MIT Bag model described by the Grand
potential \citep{Nandi:2017rhy,Nandi:2018ami}:
\begin{equation}
\Omega_{\rm QM} = \sum_i \Omega_i^0 + \frac{3\mu^4}{4\pi^2}(1-a_4)+B_{\rm eff}, \label{eq:qm}
\end{equation}
where $\Omega_i^0$ represents the grand potentials of
non-interacting Fermi gases of up ($u$), down ($d$) and strange ($s$) quarks and electrons. The other two
terms correspond to the strong interaction
correction and the non-perturbative QCD effects which are 
included via two effective parameters $a_4$ and $B_{\rm eff}$, with $\mu=\frac{1}{3}(\mu_u+\mu_d+\mu_s$) denoting the baryon
chemical potential of quarks.

In Fig. \ref{fig:eos} (left panel) we show the plots for the EoS 
(pressure vs. density) and the corresponding mass-radius 
diagrams for pure hadronic stars as well as HSs (having some amount of QM in them)
based on the S271v2 model.
The EoS of the HSs that include PT
from HM to QM is obtained via Maxwell construction and therefore characterized by a density jump, as can be seen from fig \ref{fig:eos} (a). PT makes the EoS softer and thereby
reduces the maximum mass (see fig \ref{fig:eos} (b)). The parameters of the bag model $B_{\rm eff}$ and
$a_4$ are chosen such that the overall EoS produce neutron
stars with $M_{\rm}>2M_\odot$. We construct the EoS, including PT for FSUGarnet, DD2, and SkI3, similarly. 

\begin{figure}
	\centering

		\includegraphics[width = 3.45in,height=2.5in]{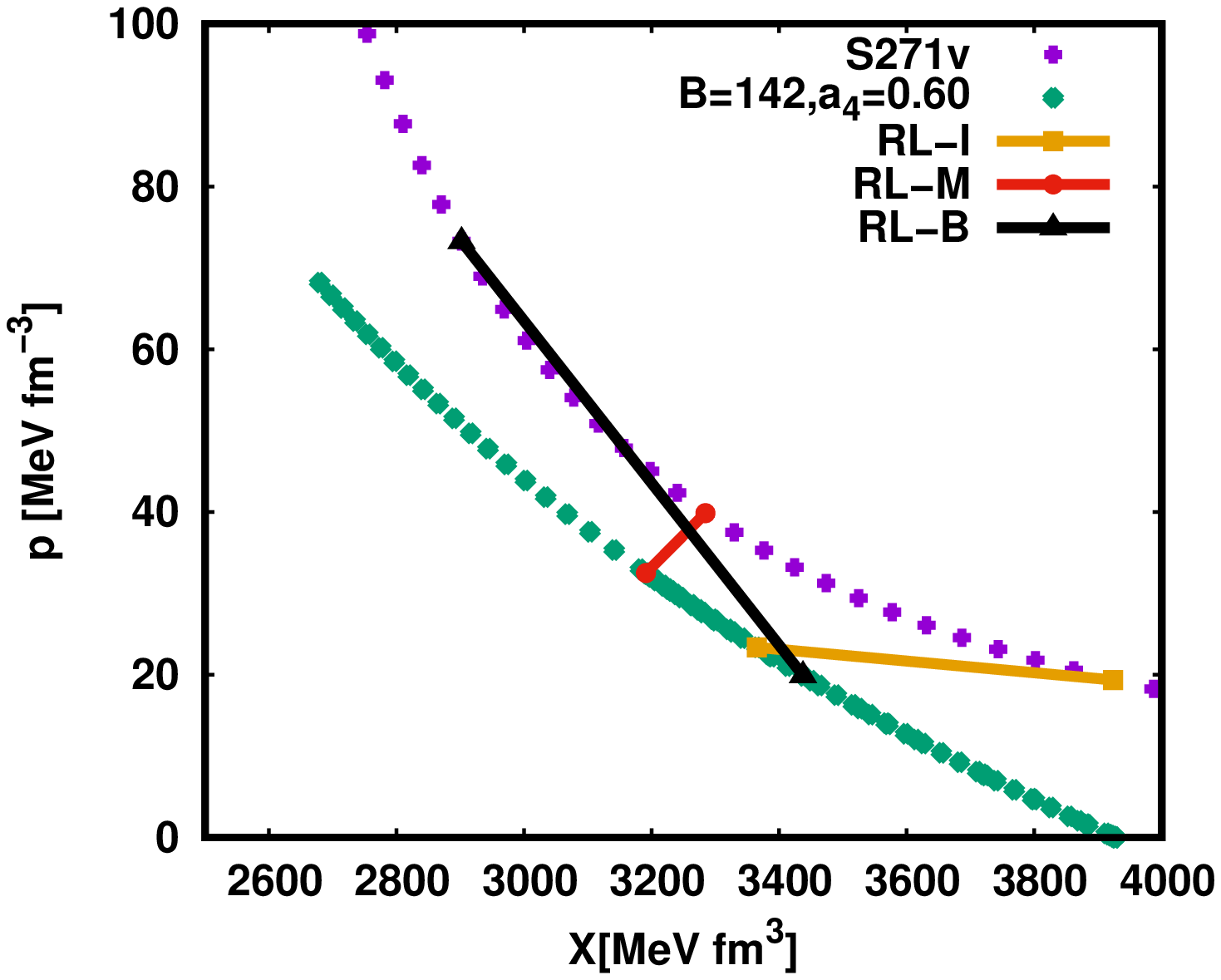}
		\includegraphics[width = 3.45in,height=2.5in]{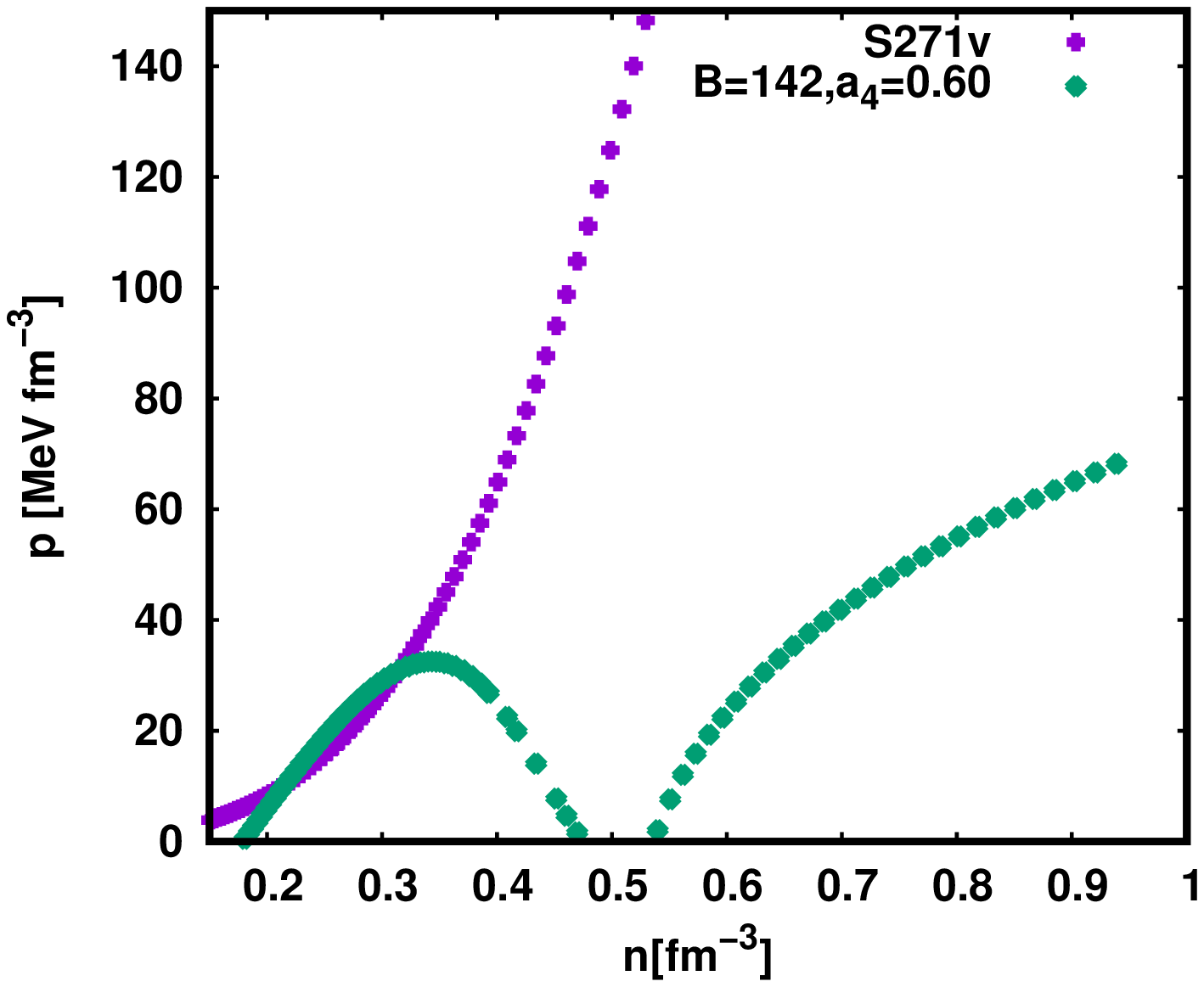}
		\hspace{0.5cm} \scriptsize{(a)} \hspace{8.5 cm} \scriptsize{(b)}

       \includegraphics[width = 3.45in,height=2.5in]{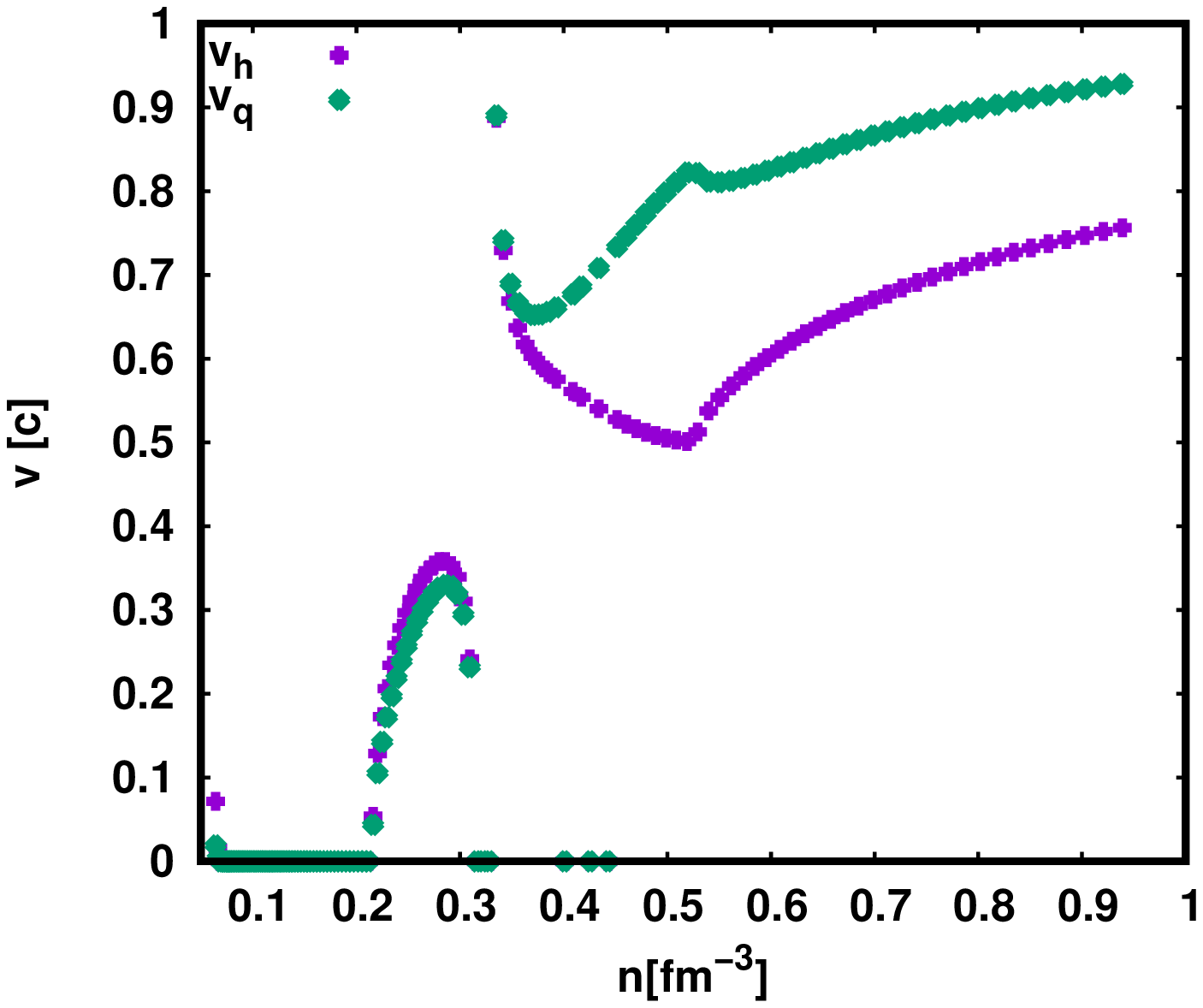}
	\hspace{8.5cm} \scriptsize{(c)}
	\caption{a) The CA (p vs. X) curves for the unburnt HM (S271v2, violet curve) and its corresponding burn state with QM ((142, 0.60), cross green curve) are drawn. The upstream point lies on the violet
curve, whereas the downstream points lie on the green curve. The RL connecting two
points of the different curve are marked as RL-I, RL-M, RL-B. The
RL connecting the downstream adiabat's maximum point with
its corresponding point on the upstream curve is marked as RL-M. The RL-I means RL connecting the matter phase before the PTLMM, and RL-B indicates the RL beyond the PTLMM.
b) $p$ as a function of $n(=n_h)$ for HM and their corresponding
downstream QM curve are illustrated in the figure (the EoS are same as that of fig \ref{fig1} (a)). The burnt pressure shows a local maximum at $n \simeq 0.35$ fm$^{-3}$ which corresponds to the RL M point of the CA curve. c) 
The upstream ($v_h$) and downstream ($v_q$) velocities are
shown as a function of n. Initially, $v_h$ is greater than $v_q$, and they both increase. As the local maximum is reached, $v_q$ becomes greater than $v_h$, and both blow up. At higher densities $v_q$ always remains greater than $v_h$.}
\label{fig1}
\end{figure}

In fig \ref{fig1} (a), we plot the CAs for the HM and QM. A particular initial configuration (a point in the HM curve) gives rise to a particular point in the QM (solving the CA equation). The whole curve is obtained by choosing different initial configurations. However, it is interesting to note that as we increase the HM pressure, the QM pressure initially rises; however, there is a maximum pressure (which can be local or a global maximum) in the burnt CA, beyond which it retraces its path even if the HM pressure increases. This is clearer in fig \ref{fig1} (b), where we find a local maximum pressure for the burnt phase even as the density increases. 

If a point in the HM (say HMC1 specifying the pressure, energy density, and density) is assumed to be the central density of the NS, we get the NS configuration (say NS1) by solving the TOV equation with the initial condition being the central density (HMC1). By solving the CA equation, we get the QM's corresponding point (say QMC1) specifying pressure, energy density, and density. Therefore, if a shock-induced PT initiates at the center of the NS1 star (the parent star), the CA equation fixes the central density of the corresponding HS (QMC1). We also get the daughter HS configuration (say HS1) by solving the TOV equation with the initial condition being the central density QMC1. Hence, each point in the CA diagram (or the pressure curve) can be used to build the parent NS configuration, and by solving the CA equation, we can also get the daughter HS, thereby mimicking the PT of an NS to HS. 

A PT occurring in a cold NS does not have any external energy source; the energy output for such a process needs to be positive since no energy source is powering the conversion. On the other hand, the maximum pressure (local or global) in the burnt phase is reflected in the HS mass. If the pressure maximum is global, we obtain a maximum mass, and if it is local, we obtain an intermediate-mass. Let us call it phase transition local maximum mass (PTLMM).
The PTLMM of the HS depends both on the quark EoS and on the hadronic EoS. The hadronic EoS gives the NS's initial configuration and solving the CA equation the HS quantities are obtained. Therefore, the initial NS EoS is also essential in determining the resultant mass of the phase transformed HS. 

We need to emphasize one thing here. The shock-induced PT is very different from equilibrium PT. In the equilibrium PT, one assumes that the HM converts smoothly to QM once the critical density (obtained by the point of crossing of the HM and QM curves in the pressure vs. chemical potential plane) is breached, for which three flavour (3f) QM is the stable ground state of matter at high density. However, we assume that the HM can exist for a short but finite time even if the critical density is breached in the shock-induced PT. For the PT to happen, a seeding has to be there (seed of stable 3-f QM). The seeding takes finite time because the conversion of two-flavor (2f)-3f QM is a weak process and is slow (although the deconfinement (strong process) is fast). The seeding gives rise to a shock discontinuity (at the boundary of NM and 3f QM), which then propagates across the star converting NM to stable 3f QM. 

The total energy output in the PT of a NS to HS is given by the difference between the total binding energy of the HS (BE(HS)) and the total binding energy of the NS (BE(NS)) \citep{ritam-npa} 
\begin{equation}
 E_T = BE(HS) - BE(NS),
\end{equation}
which is the sum of the gravitational and internal energy change 
\begin{equation}
E_T = E_I  +  E_G.  
\end{equation}

We start with the initial point (energy density) of the HM for which we get maximum pressure for the QM; we choose it to be the parent NS's central energy density. The TOV equation is solved to obtain the star configuration. The density corresponding to the local or the global maximum pressure is chosen to be the HS's central density, and solving the TOV equation, we get the HS configuration. Therefore, we get the PTLMM of the daughter HS for a chosen combination of quark and hadronic EoSs and the initial mass of the parent NS from which the daughter HS originates. If the energy output is positive, we can conclude that such a PT is possible in a cold NS. 

The total, internal and gravitational energy of conversion can be written in terms of their respective 
gravitational mass
\begin{eqnarray}
E_T = [M_G(NS) - M_G(HS)]c^2,
\end{eqnarray}
where $M_G$ is the gravitational mass of the star.

\section{Results}

\begin{table}
\centering
\begin{tabular}{cccccccc}
\hline 
EoS + (Bag, $a_4$)& $n_h$ & $n_q$& $M_{\rm max}$ & $M_{PNS}$ & $M_{\rm PTLMM}$ & $E_T (10^{53} ergs)$&\\
\hline\hline
S271v2               &&& 2.34 & \\
S271v2 + (142, 0.60) &0.110&0.243& 1.971 & 1.362 & 1.107 & 4.56 \\
S271v2 + (141, 0.55) &0.231&0.266& 2.004 & 1.924 & 1.954 & -0.529 \\
\hline
SkI3 & &&2.24 & \\
SkI3   + (142, 0.57) &0.193&0.259& 1.98 & 1.445& 1.25 & 3.48\\
SkI3   + (141, 0.55) &0.227&0.267& 2.01 & 1.922 &1.978 & -0.999\\
\hline
FSUGarnet &&& 2.067 &\\
FSUGarnet + (142, 0.60) &0.150&0.245& 1.971& 1.375 & 1.195 & 3.21\\
FSUGarnet + (141, 0.55) &0.765&0.820& 2.049& 1.981 &2.047&-1.118 \\
\hline 
DD2 &&& 2.42 & \\
DD2 + (142, 0.57) & 0.248 & 0.275 & 1.98 &1.136 &1.357 &-4.69\\
DD2 + (141, 0.55) & 0.386 & 0.405 & 2.048 &1.918 &1.944 &-0.932\\
\hline 
\end{tabular}
\caption{Table showing the maximum mass (M$_{max}$) of the sequence (row 2, colum 4), the resultant mass of the combusted HS (M$_{PTLMM}$) (column 6) and the corresponding mass of the parent NS (M$_{PNS}$) (column 5) in units of solar mass. The first column gives the EoS, the second and third colum give the central densities of the parent NS and daughter HS (in units of $fm^{-3}$). Also shown in the table the total energy release during the PT process in units of $10^{53}$ ergs (column 7).}
\label{table1}
\end{table}

Using the EoSs and following the CA prescription defined in section 2, we find the maximum pressure and thereby the PTLMM of the HS. We also calculate the corresponding mass of the parent NS and find the energy output during the PT. Table \ref{table1} shows our results. For a particular HM EoS, the 1st row (column 4) gives the NS sequence's maximum mass. The Table also gives the maximum mass of the HS sequence (with equilibrium PT, column 4). The 6th column gives the PTLMM obtained from the combustion of a cold NS to HS. The Table also shows the net energy output from the transition of a parent NS to PTLMM HS. Positive energy output indicates that the PT is possible without the supply of any energy. It also shows that not all PT results in a positive energy output, which indicates that PTLMM of the HS is not always possible through the conversion of a cold NS. It is worth mentioning that the EoSs with QM values $B_{\rm eff}^{1/4}= 141$ and $a_4=0.55$ always results in a global maximum in the intermediate density range, which indicates that that the PTLMM is the PT maximum mass for cold NSs. For the EoSs combination with QM values $B_{\rm eff}^{1/4}= 142$ ($a_4$ is either $0.57$ or $0.6$), we always get a local maximum, which suggests that the PTLMM is not the maximum mass after the PT of cold NSs.

\begin{figure*}
\centering
\includegraphics[width = 3.45in,height=2.5in]{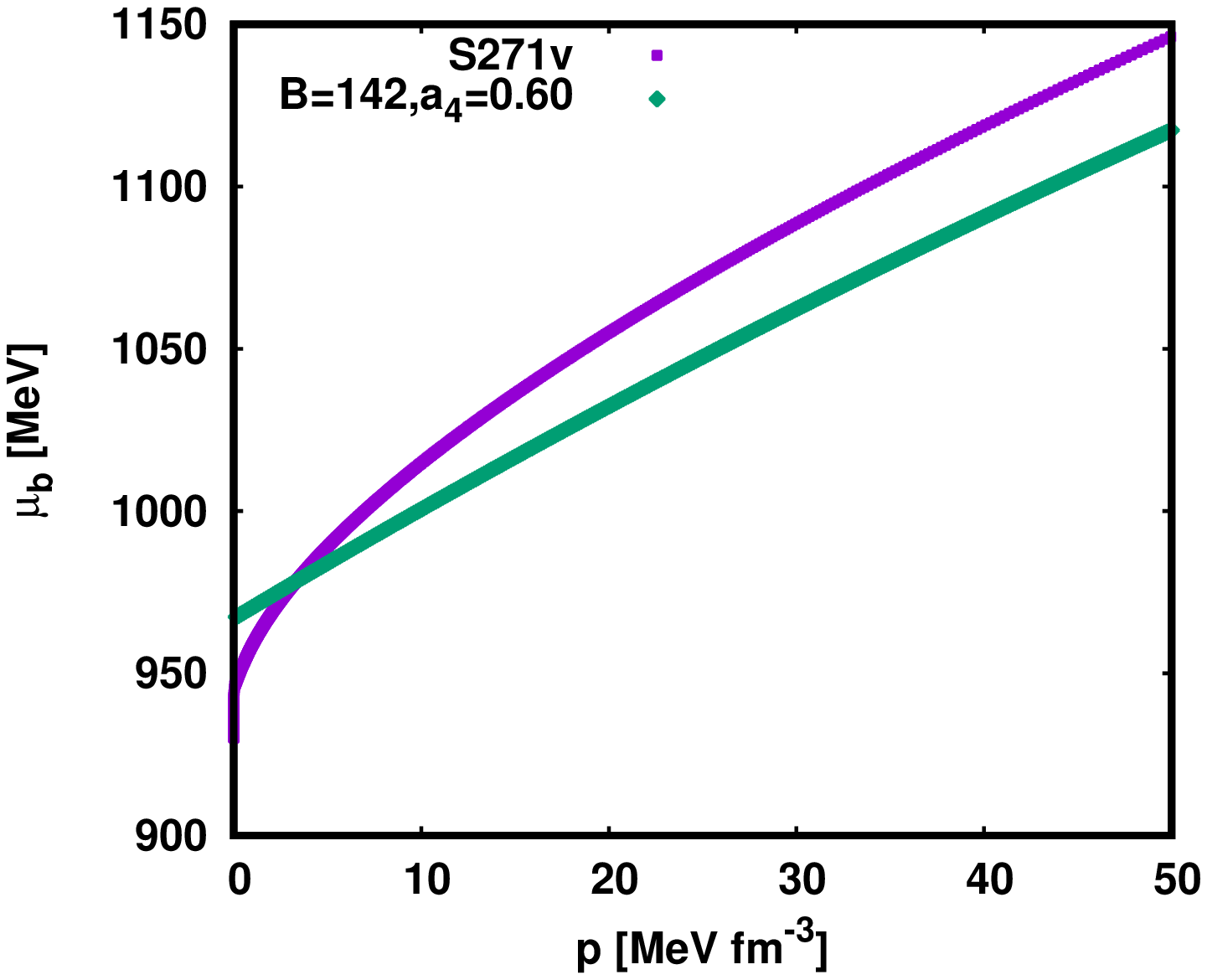}
\includegraphics[width = 3.45in,height=2.5in]{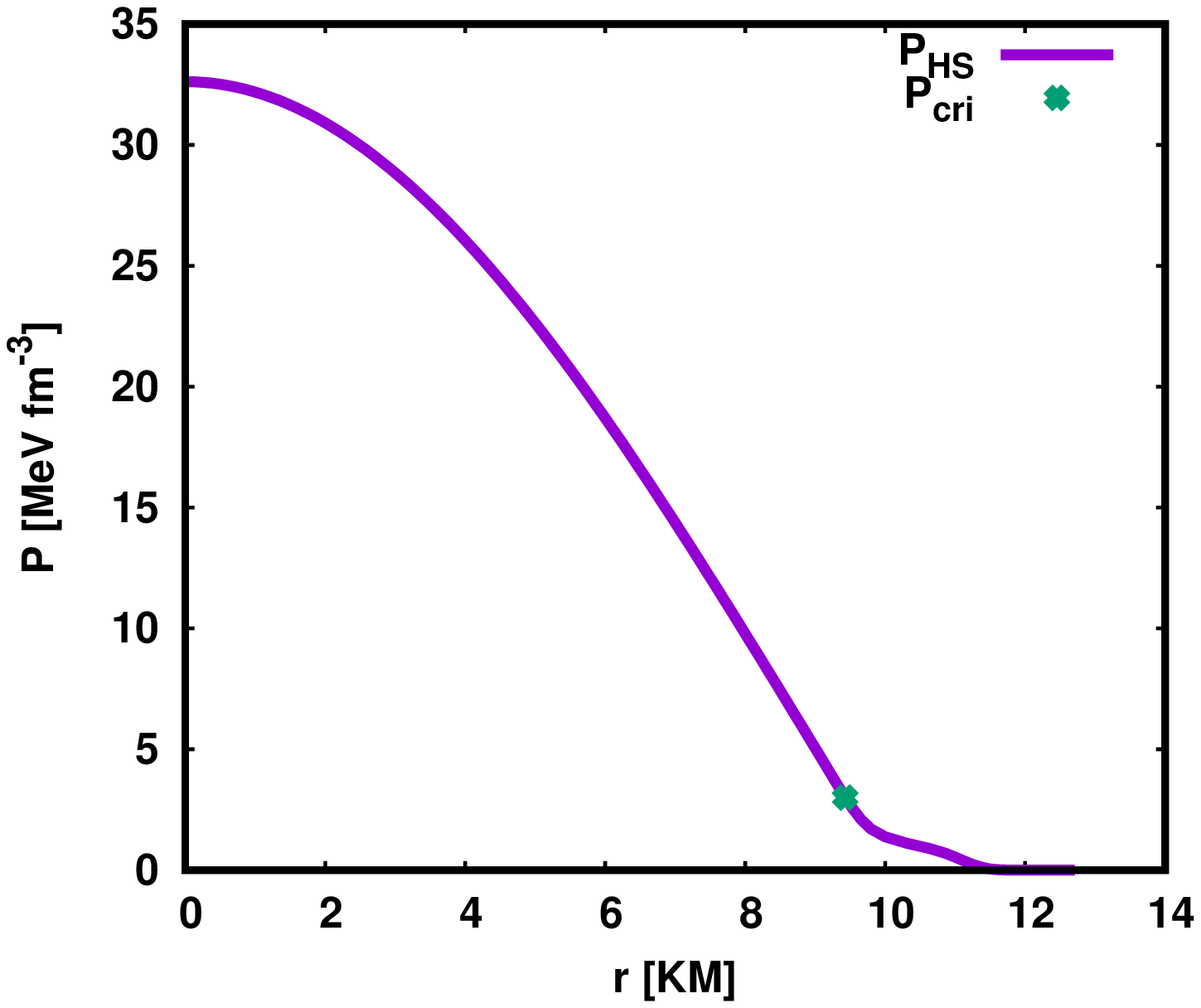}
\hspace{0.5cm} \scriptsize{(a)} \hspace{8.5 cm} \scriptsize{(b)} 
\caption{a) The $\mu_b$ vs. $p$ curves for the HM and the QM is shown in the figure. The state with lower chemical potential at a fixed pressure is the more stable state. Therefore, we find that at lower density HM is the favourable state, whereas for higher density QM is the favourable one. The two curve cross each other at $p=2.2$ MeV/fm$^3$. b) Pressure of the HS (solved using TOV) as a function of the radius is drawn. The critical density is marked by the green dot corresponding to the cross-over point of $\mu_b$ vs. $p$ curves of the left panel. At the centre of the star therefore we have QM and beyond $8.6$ km we have HM.}
\label{quark-core}
\end{figure*}

The Table also shows that if the energy output is large, then the daughter HS has a significantly smaller local maximum mass than the parent NS. As the mass difference is large, therefore a significant amount of energy is released in such PT. The PT is more likely to be observed in gravitational waves, electromagnetic spectrum, and neutrino beaming as a massive amount of neutrinos are likely to be released during PT. The noticeable fact is that for DD2 EoS, the PT to HS does not yield positive energy (for the bag parameters chosen). Therefore, the PT of a cold NS, resulting in a PTLMM, is prohibited for DD2 EoS. For the EoSs that we have considered in this calculation, even small PTLMM HS has quark cores at their center (fig \ref{quark-core}), but this may be not true for all EoSs combinations. For some EoS combination (like DD2), the exothermic PT in NS is prohibited, and external energy is needed for shock-induced PT.

We take a closer look at the matter properties for stars for which the PTLMM results in positive energy. The related matter properties for the NS and HS are shown in Table \ref{table2}. The Table shows that the quark pressure ratio (at the center of the star) to hadronic pressure and the ratio of the quark energy density to hadronic energy density is less than $1$. However, the ratio of matter velocities (quark phase velocity to hadronic phase velocity) at the star core is greater than $1$. 
In Fig \ref{fig1} (a), we plot the CA curves for the given initial state (hadronic EoS) and the burnt final state (quark EoS). Also shown in the plots are the different RLs. The pressure against the density curve is shown in fig \ref{fig1} (b). We see that a local maximum is developed around density $0.35$ fm$^{-3}$. The pressure then goes down, but later the quark pressure rises again with density; however, at the much higher pressure, thus the HS produced as such high densities are sufficiently large, but the PT process becomes endothermic. This can be seen from the RL marked in bold in fig \ref{fig1} (a). The combustion adiabat curve initially reaches a maximum (local maximum), then goes down again for some time, and then again rises. This is a typical behavior noticed in the PTLMM of HS for which the reaction process is exothermic. However, as the pressure maximum happens at a lower density, the star's mass is not very large. 
On the other hand, the HS can be significantly massive at higher densities; however, the PT process in which the quark core is generated has to be endothermic.

The condition for detonation and deflagration is given by 
\begin{align}
 v_q \ge v_h \quad \text{detonation  } \nonumber \\
 v_q \le v_h \quad \text{deflagration}
\end{align}

If we assume that the PT starts from the star center and spreads spherically to the surface, we find that to have positive energy output or to have a PT resulting in a PTLMM; the conversion should start as a detonation (fig. \ref{fig1} (c)). 
The velocity diagram (fig \ref{fig1} (c)) shows that at the point of local maximum pressure, the burnt phase matter velocity becomes greater than the unburnt phase matter velocity indicating that the deflagration turns into a detonation. Therefore, for HS smaller than PTLMM, the PT at the core starts as a deflagration; however, the PTLMM's PT process initiation at the star center is a detonation. At the point for which the PTLMM is obtained, the matter velocities blow up. Before this point we have $v_h > v_q$ indicating deflagration and beyond the point one has $v_q> v_h$ indicating detonation. 
A detonation is needed for the PT For massive stars; however, energy must be supplied.

\begin{table}
\centering
\begin{tabular}{cccccccccc}
\hline 
EoS&$p_q$ & $\epsilon_q$&$E_T$&$\frac{p_q}{p_n}$&$\frac{\epsilon_q}{\epsilon_n}$&$\frac{v_q}{v_h}$\\
\hline\hline
S271v2 + (142, 0.60)&32.50&330.25&4.56&0.82&0.96&1.02\\
\hline
SkI3   + (142, 0.57)&41.86&359.90&3.48&0.88&0.95&1.03\\
\hline
FSUGarnet + (142, 0.60)&42.02&359.74&3.2&0.89&0.95&1.033\\
\hline
\end{tabular}
\caption{Table showing the pressure, energy density of QM which gives positive energy at the PTLMM. Also shown in the table the pressure, energy density and velocity ratios of the two phases along with the total energy output.}
\label{table2}
\end{table}

Having studied the properties of the matter quantities for which we get a PTLMM with positive energy output, we next study the properties for which the energy output is negative.
In Table \ref{table3}, we show the corresponding matter properties for which we get a negative energy output. The Table shows that the quark pressure ratio (at the center of the star) to hadronic pressure and the ratio of the quark energy density to hadronic energy density is greater than $1$. Also, the ratio of matter velocities (quark phase velocity to hadronic phase velocity) at the star core is less than $1$. That means at the core of the star; the conversion process is a deflagration. Such a conversion process is not tenable unless some external energy source supplies energy to the parent NS.

\begin{table}
\centering
\begin{tabular}{cccccccccc}
\hline 
EoS&$p_q$ & $\epsilon_q$ &$E_T$&$\frac{p_q}{p_n}$&$\frac{\epsilon_q}{\epsilon_n}$&$\frac{v_q}{v_h}$\\
\hline\hline
S271v2 + (141, 0.55)&160.21&725.72&-0.529&1.53&1.46&0.79\\
\hline
SkI3  + (141, 0.55)&187.31&808.56&-0.999&1.49&1.4&0.82\\
\hline
FSUGarnet + (141, 0.55)&301.44&1157.34&-1.117&1.82&1.65&0.76\\
\hline
DD2 + (141, 0.55)&156.73&714.99&-4.67&1.56&1.51& 0.77\\
\hline
\end{tabular}
\caption{Table showing the pressure, energy density of QM which gives negative energy at the PTMM. Also shown in the table the pressure, energy density and velocity ratios of the two phases along with the total energy output.}
\label{table3}
\end{table}

\begin{figure}
	
		\centering
		\includegraphics[width = 3.45in,height=2.5in]{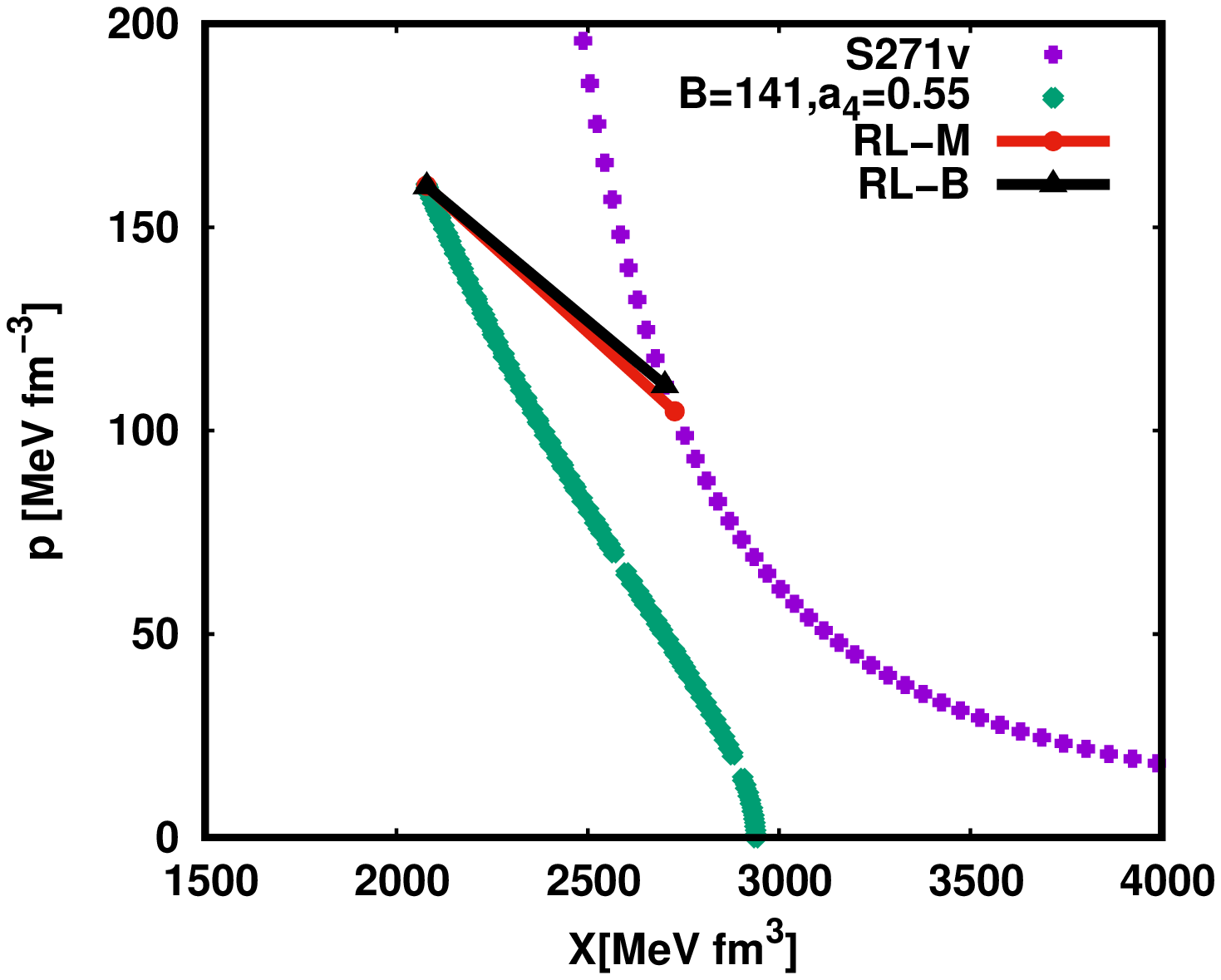}
		\includegraphics[width = 3.45in,height=2.5in]{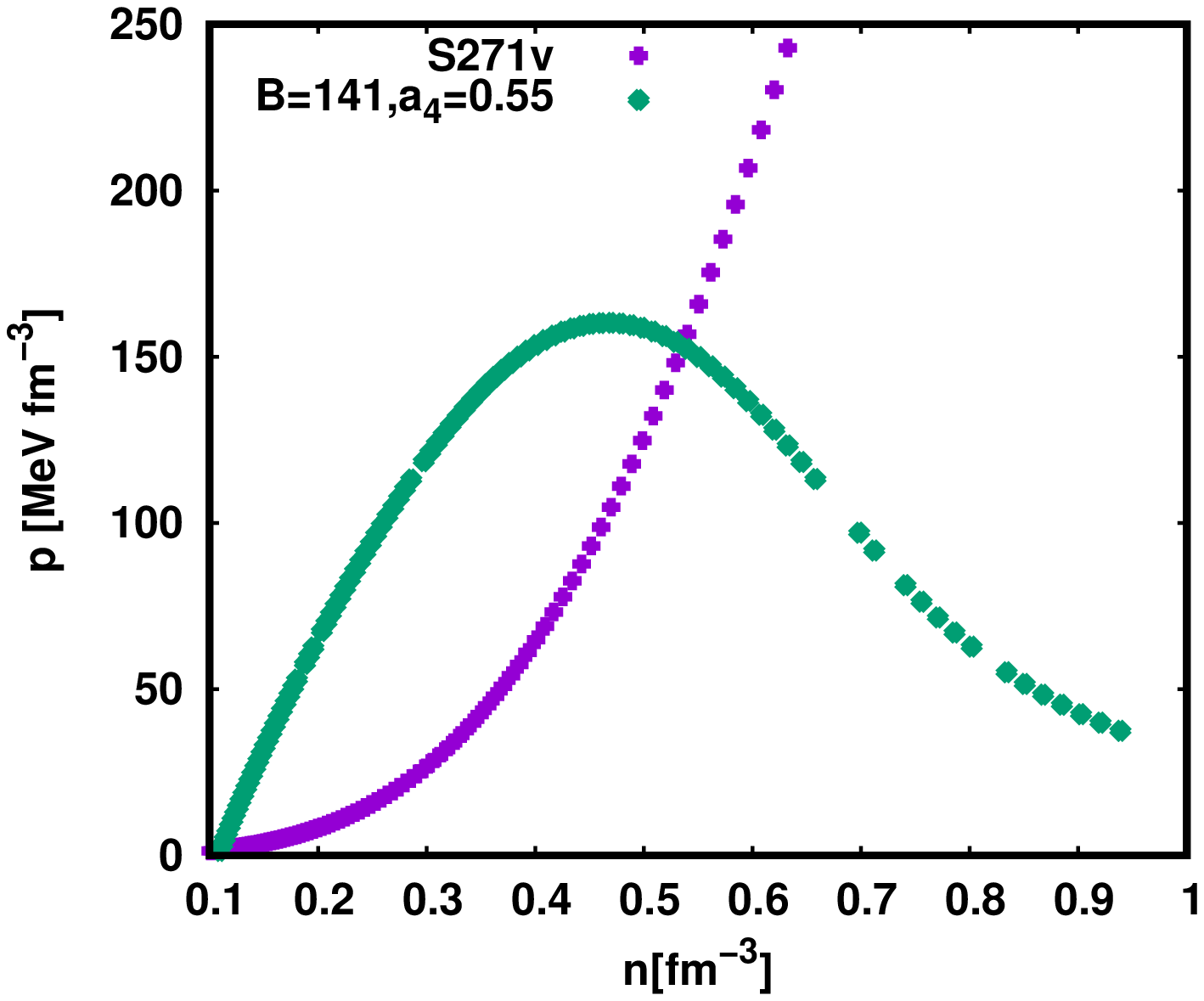}
			\hspace{0.5cm} \scriptsize{(a)} \hspace{8.5 cm} \scriptsize{(b)}
    	\includegraphics[width = 3.45in,height=2.5in]{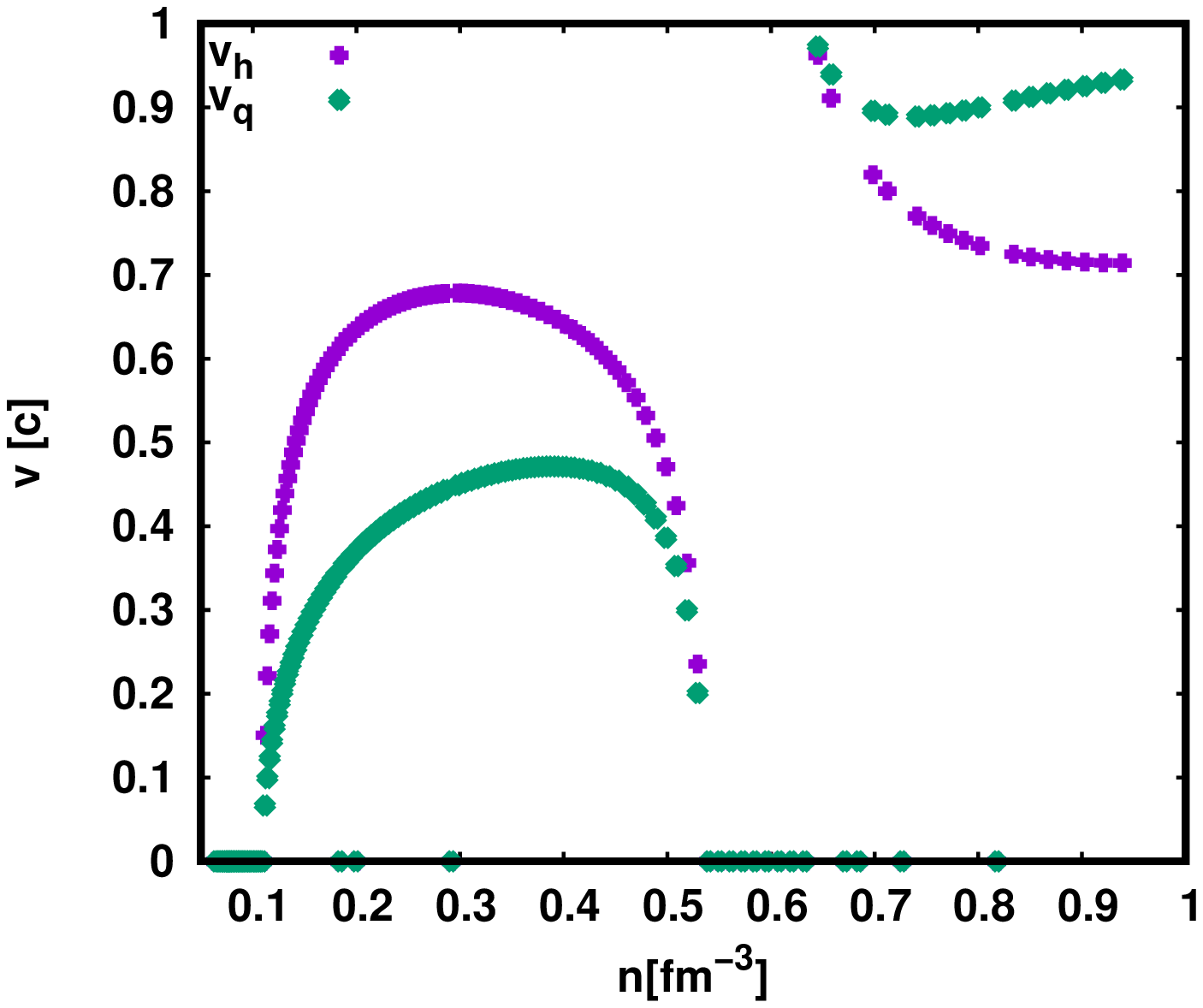}
	  \hspace{8.5cm} \scriptsize{(c)}
\caption{a) The CA (p vs. X) curves for unburnt HM (S271v2)  and its corresponding burn
state with QM curve (141, 0.55) are drawn. This EoS combination results in a negative energy output at PTMM. 
The upstream point lies on the HM curve, whereas the downstream points lie on the QM. The
RL are the straight lines. Other nomenclature remains the same as fig \ref{fig1}.
b) $p$ as a function of $n$ for HM and their corresponding
downstream QM curve are illustrated in the figure (EoS are same as fig \ref{fig3} (a)). The burnt pressure shows a global maxima at $n=0.47$ fm$^{-3}$ which corresponds to the RL-M point of the CA curve. This results in the PTLMM being the true PTMM.
c) The upstream ($v_h$) and downstream ($v_q$) velocities are
shown as a function of n. Initially, $v_h$ is greater than $v_q$, and they both increase. As the local maxima are reached, they attain a maximum point and then goes to zero at $n=0.55$ fm$^{-3}$. At higher densities, both again becomes non zero, and the curve nature changes drastically with $v_q$ always remaining greater than $v_h$.}
\label{fig3}
\end{figure}

Looking at the CA diagram (fig \ref{fig3} (a)), we see that the maximum pressure at the intermediate region is not a local maximum but a global one. This is the true maxima, as shown by the RL. In this case, the burnt phase pressure rises with the hadronic phase pressure and finally reaches a maximum. It then retraces its path and goes down. This is also seen clearly from the pressure diagram (fig \ref{fig3} (b)), where the pressure at some intermediate density is the maximum. Therefore, the PTLMM is the phase transition maximum mass (PTMM). 
For these EoSs, the negative energy output does not necessarily mean that PT is prohibited for such a star sequence. It only means that PT cannot happen (unless external energy is supplied) to obtain the PTMM of the HS sequence. However, for significantly higher densities (at the core of parent NSs), the PT can result in positive energy. In table \ref{table4} we show this result. Here the central density of the parent NS is higher than the central density for which we have PTMM. We find that the energy for such a conversion process is positive. The Taub adiabat diagram, fig \ref{fig3}, shows how the energy from being negative becomes positive. Initially, for the PTMM star, the RL slope is very stiff, which indicates a large difference in the central pressure of the initial parent star and the final daughter star. An external energy source is needed to sustain such a process, however, as the hadronic pressure increases (mass of NS also increases), the combustion adiabat retraces its path, and the burnt pressure QM reduces. The RL slope reduces after a certain point the pressure difference between the parent NS and daughter HS becomes such that no external energy is needed to sustain such pressure difference, and the PT can result in positive energy. This means a massive NS undergoes a PT and results in a massive HS. The masses of the parent NS and daughter HS are shown in Table 4. The difference in the masses is small, and therefore the energy output is also small. 

From the velocity diagram (fig \ref{fig3} (c)), we see that mostly the QM velocity is smaller than HM velocity, resulting in a deflagration process at the star center. For most of the stars, the PT at the star core starts as a deflagration process. For massive stars, the PT starts as a detonation. We should emphasize that for some density range, the matter velocity comes out to be imaginary; however, for practical purposes, we have assumed them to be zero (the zeros in the velocity diagram).

\begin{table}
\centering
\begin{tabular}{cccccccccccc}
\hline 
EoS&$p_q$ & $\epsilon_q$&$E_T$&$\frac{p_q}{p_n}$&$\frac{\epsilon_q}{\epsilon_n}$&$\frac{v_q}{v_h}$&$M_{PNS}$&$M_{DHS}$\\
\hline\hline
S271v2 + (141, 0.55)&159.99&725.06&0.22&1.44&1.42&0.8&1.955&1.953\\
\hline
SkI3  + (141, 0.55)&184.9&801.17&0.27&1.23&1.28&0.86&1.99&1.975\\
\hline
FSUGarnet + (141, 0.55)&175.93&844.02&0.01&1.02&1.18&0.88&2.053&2.04\\
\hline
DD2 + (141, 0.55)&156.42&714.14&0.006&1.42&1.45& 0.79&1.976&1.973\\
\hline
\end{tabular}
\caption{Table showing the pressure, energy density for densities beyond PTMM for EoSs which gives negative energy at PTLMM. It shows that for higher dnsities (corresponding to higher pressure and energy density) the energy output is positive. Also shown in the table the pressure, energy density and velocity ratios of the two phases along with the total energy output. The masses of the parent NS and daughter HS (M$_{DHS}$) is also shown.}
\label{table4}
\end{table}

\section{summary and discussion}

In this study, we have investigated the PT process starting at the core of a cold NS via a shock. Different HM and QM EoS are used in our analysis. We have chosen the EoS combination for which the NS and the HS are consistent with the recent astrophysical and nuclear bounds. Using the CA technique, we find the global or the local maximum pressure, thereby the PTMM or a PTLMM of the HS and the parent NS's corresponding mass. We also find the energy output of the PT process. For some EoS combinations, we find that the PT process results in positive energy (PTLMM); however, for some EoS combinations, the PT process at PTMM is endothermic.

We find that the EoS combination for which the energy output is large and positive, the resulting HS has a significantly smaller mass than the parent NS. As the mass difference is substantial, a significant amount of energy is released in such PT. Such a PT process is likely to be observed through gravitational waves and neutrino beaming due to generations of large numbers of neutrinos in the PT process. In such stars, the onset of the hadron to quark conversion happens at very low density, and even relatively small stars contain a quark core.
To have a positive energy output with a PTLMM, the star core conversion starts as a detonation. For PTLMM,
the combustion adiabat curve initially reaches a maximum (local maximum), then goes down for some time, and then again rises. This is a typical behavior noticed for such EoS combinations where the star has a PTLMM, and the reaction process is exothermic. 
However, as it happens at a lower density, the star is not massive.
At much higher densities, the pressure again rises, and we can have significantly massive HSs.
However, for such massive stars, the PT process in which the quark core is generated has to be endothermic. This can happen during the supernova or in a hypermassive NS formed by binary NS collision. The combustion process at these densities is detonation and needs an external energy source to initiate the process. 

The HSs, which are smaller than the PTLMM, the PT at the core starts as a deflagration, and the energy output is positive. This is because the star cores' pressure difference (between the parent NS and daughter HS) is high. As the NS mass increases (increase in the central pressure), the pressure difference reduces, thereby reducing the mass difference between the parent and daughter star, and the energy output also reduces. At the PTLMM, the pressure difference is such that the combustion process at the star core changes from deflagration to detonation; however, it is still exothermic. For more massive stars (having higher central pressure), the pressure difference reduces, and the combustion process remains a detonation, but the process becomes endothermic.

The star core's combustion process starts as a deflagration for EoSs combination that results in a PTMM, but with negative energy output. Such a conversion process is not tenable unless some external energy source supplies energy to the parent star. The pressure for obtaining the PTMM is a global maximum pressure. In this case, the burnt phase pressure rises with the hadronic phase pressure and finally reaches a maximum. 
For much massive parent NSs (having higher central pressure), the pressure difference between the parent NS and daughter HS reduces, and the PT process can become positive, and the mass of daughter hybrid star is quite massive. 

Our analysis shows that PT happening in a cold NS can be both exothermic and endothermic. The exothermic PT can result in significant energy release, which can have observational significance. For massive stars to have QM at their center, the PT usually is endothermic, which means such stars develop their quark core during the star-forming phase of supernova explosion or hypermassive star after binary NS merger. However, for a few cases, massive HS can also result from exothermic PT.

\section{acknowledgements}
The author RM is grateful to the SERB, Govt. of India, for monetary support in the form of Ramanujan Fellowship (SB/S2/RJN-061/2015).  RM and SS would also like to thank IISER Bhopal for providing all the research and infrastructure facilities.

\end{document}